\documentclass[12pt]{article}
\usepackage{amsmath}
\usepackage{amsfonts}
\usepackage{appendix}
\begin{document}
\title{Massless Fermions in planar Bianchi-type-I universes: Exact and approximate solutions}
\author{Matthias Wollensak\footnote{matthias.wollensak@uni-jena.de} \\Theoretisch-Physikalisches Institut,
\\Friedrich-Schiller-Universit$\ddot{\mathrm{a}}$t Jena,
\\Max-Wien-Platz 1, D-07743 Jena, Germany}
\date{13. June 2021}
\maketitle

\begin{abstract}
Based upon the exact formal solutions of the Weyl-Dirac-equation in anisotropic planar Bianchi-type-I background spacetimes with power law scale factors, one can introduce suitable equivalence classes of the solutions of these models. The associated background spacetimes are characterized by two parameters. It is shown that the exact solutions of all models of a given equivalence class can be generated with the help of a special transformation of these two parameters, provided one knows a single exact solution of an arbitrary member of this class.

The method can also be utilized to derive approximate solutions, i.e. solutions which exhibit the correct behavior at early and at late times as well. This is explicitly demonstrated for the case of the anisotropic Kasner background with axial symmetry.
\end{abstract}

\section*{I. INTRODUCTION}
The study of particles, which obey Dirac's equation and propagate in curved spacetime backgrounds,  has occupied physicists for quite some time because of their importance in cosmological problems. An early discussion of the massless case in spherically symmetrical backgrounds has been given by Brill and Wheeler \cite{Brill}. Quantized spin-$\frac{1}{2}$-fields in (conformally) flat Friedmann-Lemaitre-Robertson-Walker (FLRW) universes have been investigated by Parker, who showed that there is no production of massless particles in those spacetimes \cite{Parker}.

An incentive for the study of (quantized) fields in anisotropic spacetimes is given by the problem that, although the present day universe appears to be highly isotropic, this needed not necessarily be the case in a very early phase of development of the universe. In fact, if one starts with an anisotropically expanding Kasner universe and takes into account quantum effects in the vicinity of the initial singularity, then, following Zel`dovich  \cite{Zel`dovich2} this could lead to an isotropization of the universe at the Planck time-scale due to particle creation processes. Kasner spacetimes represent a subclass of Bianchi-type-I (BI) spacetimes. These describe anisotropic homogeneous universes which are spatially flat. BI universes exhibit along each spatial axis a different time-dependent scale factor and are a natural generalization of flat FLRW universes. The special Kasner case is characterized by power law scale factors, whose exponents satisfy the well-known Kasner conditions. Since the Kasner metric is an exact vacuum solution of Einstein's field equations, it is suitable to approximate the situation of a very early universe, where matter terms in the field equations are small in comparison with typical nonzero entries of the Riemann tensor. A semiclassical calculation carried out by Hu and Parker \cite{Hu} lent further credit to Zel`dovich's point of view. In their calculation they investigated a quantized massless conformal scalar field in a BI universe with axial symmetry.

A further motivation for the special interest in Kasner universes as background spacetimes is given by the work of Belinskii, Khalatnikov, and Lifshitz \cite{BKL} and Misner \cite{Misner}, who realized that Bianchi-type-IX universes can be described by sequences of Kasner spacetimes, if one moves backward in time toward the initial singularity.

More recently, the issue of anisotropic BI spacetimes has been discussed in the context of preinflationary scenarios of the universe. One takes e.g. as background geometry a general BI spacetime, which near the singularity and before the onset of inflation assumes the form of a Kasner spacetime \cite{Pitrou} - \cite{Kim}. Anisotropy also occurs in vector inflation models. To establish an isotropic FLRW background during cosmic inflation, one therefore introduces, instead of considering only a single vector field, either three mutually orthogonal vector fields with the additional constraint that these fields must have exactly the same magnitude ("cosmic triad") \cite{Armendariz}, \cite{Golovnev}, or alternatively a large number of randomly oriented vector fields \cite{Golovnev}. However, in the latter case a small amount of a few percent of global anisotropy survives the inflationary phase. Moreover, in the dark energy era there is no need for isotropization and it stands to reason to employ in this case a BI spacetime as background. The use of such a spacetime also allows the relaxation of the above constraint on the cosmic triad \cite{Mota}. BI spacetimes appear in anisotropic gauge inflation, too. In particular, there exist exact solutions with backgrounds described by planar BI spacetimes with power-law scale factors \cite{Malaknejad}.

At least in some cases it is relatively easy to find exact massive classical solutions of Dirac`s equation in isotropic backgrounds, which is no longer true for anisotropic backgrounds. For example, Barut and Duru investigated massless and massive fermions in two flat FLRW spacetimes with power-law expansion, and in the steady-state part of de Sitter spacetime \cite{Barut}. The latter model had also been studied by Cotaescu, who in addition performed the quantization of the spin-$\frac{1}{2}$-field \cite{Cotaescu}, and by Candelas and Raine using a path integral approach for the quantization of massive scalar and Dirac fields \cite{Candelas1}. Massless and massive fermion propagators in flat FLRW backgrounds with constant deceleration have been treated by Koksma and Prokopec \cite{Koksma}. The time-dependent fermion mass in their model is generated by a scalar field. For analytical treatment, this field must be proportional to the Hubble parameter.

To obtain analytical results when quantizing fermionic fields propagating in anisotropic spacetimes one usually has to resort to a  perturbative treatment of the background. One considers e.g. in a method utilized by Zel`dovich and Starobinsky small anisotropic perturbations about a flat FLRW spacetime \cite{Zel`dovich1}. For a model with a very special form of  weak anisotropy, Birrell and Davies performed the quantization of the massive scalar field \cite{Birrell}, and Lotze treated the quantization of the corresponding massive spin-$\frac{1}{2}$-field \cite{Lotze}.

Classical solutions of the Dirac equation in BI spacetimes have been scrutinized for example by Henneaux who considered gravitational and spinor-fields being both invariant under a special group of transformations \cite{Henneaux}, or by Saha and Boyadjiev who studied (interacting though space-independent) spinor and scalar fields \cite{Saha}.

Despite some efforts, however, even in the massless case only few exact results describing spin-$\frac{1}{2}$-particles in anisotropic backgrounds are known. Moreover, it is by no means trivial to obtain at least approximate solutions. These have the desirable property to match the exact solutions at early and late times. Recently, in order to obtain such approximate solutions, it has been proposed to utilize a time-evolution-operator (TEO) approach \cite{Wollensak1}. In that work the approximate Weyl-Dirac-TEO has been calculated for planar BI backgrounds with power law scale factors (pBI).

It is shown in this paper that, upon introducing equivalence classes $\boldsymbol{ [\,\delta \,]}$ of pBI spacetimes and related equivalence classes $\boldsymbol{[\,\bar{\delta}\,]}$ of exact spinor solutions in those pBI backgrounds, all solutions belonging to a given class $\boldsymbol{[\,\bar{\delta}\,]}$ can be obtained by a simple parameter transformation, provided one knows an arbitrary member of $\boldsymbol{[\,\bar{\delta}\,]}$. In sect. III we derive this exact transformation, and in sect. IV exactly solvable equivalence classes are presented, while sect. V deals with approximate solutions in the above sense. As an example the determination of all approximate solutions of the class $\boldsymbol{[\,\overline{1/4}\,]}$ will be performed, which in particular includes the case of Weyl-spinors in the presence of the axisymmetrical Kasner background with Kasner exponents ($ \frac{2}{3}, \frac{2}{3}, - \frac{1}{3}$).

\section*{II. BASIC EQUATIONS}
Dirac`s equation in a BI background spacetime reads w.r.t. an orthonormal frame \cite{Wollensak1}:

\begin{equation}
\label{Gl2}
(\gamma^\mu D_{e_\mu} + i m ) \, \psi \, \equiv  \,  \gamma_0  \left( e_0 - \gamma_0 \sum\limits_{j=1}^{3}
\gamma_j e_j  +  \mathcal{C} + i \gamma_0 m \right) \psi = 0   ,
\end{equation}

with $D$ the covariant differential, $\mathcal{C} := \partial_t \, $ln$|g|^{1/4}$, and line element

\begin{equation}
\label{Gl1}
ds^2 = \eta_{\mu \nu} \, \Theta^\mu \Theta^\nu
\end{equation}

$(\eta_{\mu \nu} =$ diag(1, -1, -1, -1)). The basis vectorfields are defined by $e_0 = \partial_t, \ e_j = \alpha_j^{- 1}(t) \partial_j$, and the corresponding covectorfields (1-forms) are: $\Theta^0 = dt, \ \Theta^k = \alpha_k(t) dx^k$. Vector - and covectorfields satisfy: $\Theta^\nu(e_\mu) = \delta^\nu_\mu$. The metric tensor is $\boldsymbol{g} = \eta_{\mu\nu} \Theta^{\mu} \otimes \Theta^{\nu} \equiv g_{\mu \nu} dx^\mu \otimes dx^\nu , \ g_{\mu \nu} =$ diag(1, -$\alpha_1^2$, -$\alpha_2^2$, -$\alpha_3^2$)  and $g =$ det$g_{\mu \nu}$. Spatial translational invariance of (\ref{Gl1}) suggests the ansatz

\begin{equation}
\label{Gl3}
\psi_\textbf{k} (\textbf{x},t) = c_\textbf{k} \, e^{i\textbf{k} \textbf{x}} \left( \begin{array}{c} \varphi(\textbf{k}, t)  \\ \chi(\textbf{k}, t)  \end{array} \right)   .
\end{equation}

In the massless case, which can be viewed as an approximation to the ultrarelativistic massive fermion case,  eq. (\ref{Gl2}) decouples by setting $\chi = \mp \varphi$. $\chi, \, \varphi$ denote Weyl-spinors, and in the following the superscript signs correspond to the two chirality eigenstates, i.e. we have with (\ref{Gl3}): $\psi^{(\mp)}_\textbf{k} = c^{(\mp)}_\textbf{k} e^{i\textbf{k} \textbf{x}}  (\varphi^{(\mp)}, \mp \varphi^{(\mp)})^T$. It is now convenient to introduce new Weyl-spinors

\begin{equation}
\label{Gl3a}
\phi^{(\mp)}(\textbf{k}, t) \, = \, \left( \begin{array}{c} \phi_1^{( \mp)}(\textbf{k}, t)  \\ \phi_2^{(\mp)}(\textbf{k}, t)  \end{array} \right)   ,
\end{equation}

whose components are related to those of the spinors $\varphi^{(\mp)}$ via:

\begin{equation}
\label{Gl4}
\varphi_J^{(\mp)}(\textbf{k}, t) \, = \,  |g(t)|^{- 1/4} \, \exp \left[ \mp (- 1)^J \, i P_3(k_3, t; t_{\widetilde{A}})  \right]  \, 
\phi_J^{(\mp)}(\textbf{k}, t)
\end{equation}

($J = 1, 2$), with

\begin{equation}
\label{Gl5}
P_3(k_3, t; t_{\widetilde{A}}) = \int \limits_{t_{\widetilde{A}}}^{t} p_3(y) dy  
\end{equation}

$(t_{\widetilde{A}} \geq 0)$, and $p_j(t) := k_j/\alpha_j(t)$ denote the components of the physical 3-momentum. Use of (\ref{Gl3a}), (\ref{Gl4}) reduces eq. (\ref{Gl2}) to:

\begin{equation}
\label{Gl6}
\partial_t \phi^{(l, \mp)} - \Omega^{(\mp)} \phi^{(l, \mp)} = 0   ,
\end{equation}

where $l$ numbers for each chirality the two independent solutions of (\ref{Gl6}), and

\begin{equation}
\label{Gl7}
\begin{aligned}
\Omega^{(\mp)}(\textbf{k}, t) \, = & \, \left( \begin{array}{rr} 0 \  \  \  \  \  \  \  \    &  \mathcal{P}^{(\mp)}(\textbf{k}, t)
\\ - [ \mathcal{P}^{(\mp)}(\textbf{k}, t) ]^\ast  &  0 \  \  \  \  \  \   \end{array}   \right)   ,
\\
\mathcal{P}^{(\mp)}(\textbf{k}, t) \, = & \, \pm (i p_1 + p_2) \, e^{\mp 2i P_3(k_3, t; t_{\widetilde{A}})}   .
\end{aligned}
\end{equation}

Provided  $\phi^{(1, \mp)} = (\phi^{(1, \mp)}_1, \phi^{(1, \mp)}_2)^T$ is an exact solution of the system (\ref{Gl6}), (\ref{Gl7}), then the second exact solution can always be chosen to be orthogonal to this solution w.r.t. the hermitean scalar product $(\varphi, \psi) = \sum_J \varphi^\ast_J \psi_J$  due to the special form of $\Omega^{(\mp)}$. Hence, $\phi^{(2, \mp)} = (\phi^{\ast \, (1, \mp)}_2, - \phi^{\ast \, (1, \mp)}_1)^T$.

On solving (\ref{Gl6}) one gets with (\ref{Gl3}) - (\ref{Gl4}) the four bispinor solutions:

\begin{equation}
\label{Gl8}
\psi_\textbf{k} ^{(l, \mp)}(\textbf{x},t) = c^{(l, \mp)}_\textbf{k} \, \frac{e^{i\textbf{k} \textbf{x}}}{|g(t)|^{\frac{1}{4}}} \,\left( \begin{array}{c} 
e^{\pm i P_3(k_3, t; t_{\widetilde{A}})} \, \phi_1^{(l, \mp)}(\textbf{k}, t) 
\\   
e^{\mp i P_3(k_3, t; t_{\widetilde{A}})} \, \phi_2^{(l, \mp)}(\textbf{k}, t)
\\
\mp \, e^{\pm i P_3(k_3, t; t_{\widetilde{A}})} \, \phi_1^{(l, \mp)}(\textbf{k}, t)
\\
\mp \,  e^{\mp i P_3(k_3, t; t_{\widetilde{A}})} \, \phi_2^{(l, \mp)}(\textbf{k}, t)
\end{array} \right) .
\end{equation}

In the ensuing calculation it suffices to consider negative chirality Weyl-spinors $\phi^{(l, -)}$, since the positive chirality spinors are given by

\begin{equation}
\label{Gl9}
\phi^{(l, +)}(\textbf{k}, t) = \phi^{(l, -)}(- \textbf{k}, t)  .
\end{equation}

For simplicity we specialize now to BI spacetimes (\ref{Gl1}) with axial symmetry. Moreover, since we are also interested in the evolution of a BI background at very early times, the scale factors can be assumed to exhibit power law behavior. This is motivated by the fact that close to the singularity a BI spacetime can be well approximated by a suitable vacuum Kasner geometry \cite{Lif}. Hence, in the following we consider background geometries described by
\begin{equation}
\label{Gl10}
ds^2 = dt^2 - t^{2 \nu} (dx^1)^2 -  t^{2 \nu} (dx^2)^2 -  t^{2 - 2 \mu} (dx^3)^2  ,
\end{equation}

and for later convenience we definine the parameter

\begin{equation}
\label{Gl11}
\delta(\mu, \nu) = (1 - \nu)/\mu
\end{equation}

($\mu \neq 0$), and the time variables

\begin{equation}
\label{Gl12}
s(t) = t^\mu, \  \  \  \ \tau(s) \, = \, 2 k_3 s/\mu  .
\end{equation}

By use of the linear (negative chirality) operator $\hat{\Omega}_{\textbf{k}}$ defined by

\begin{equation}
\hat{\Omega}_{\textbf{k}}[\xi] = \xi_A + \int \limits_{t_A}^t  \Omega^{(-)}(\textbf{k}, y) \, \xi(\textbf{k}, y) dy   \nonumber
\end{equation}

with initial condition $\xi_A \equiv \xi(\textbf{k}, t_A), \, t_A \geq t_{\widetilde{A}} \geq  0$, the problem of solving (\ref{Gl6}) can be cast into an equivalent fixed point problem: $\hat{\Omega}_{\textbf{k}}[\phi^{(l, -)}] = \phi^{(l, -)}$. It can be proven that there exists exactly one such fixed point, and that for any continuous function $\xi(\textbf{k}, t)$ holds \cite{Weissinger}:

\begin{equation}
\phi^{(l, -)}(\textbf{k}, t) \, = \, \lim_{n \to \infty} \left( \hat{\Omega}_{\textbf{k}}^n[\xi] \right) (t)   .   \nonumber
\end{equation}

As a consequence, the $exact$ formal solutions to eq.(\ref{Gl6}) can be written as:

\begin{equation}
\label{Gl13}
\phi^{(l, -)}(\textbf{k} ,t)  = K_{\textbf{k}}^{(-)}(t|t_A) \, \phi^{(l, -)}_A (\textbf{k})  ,
\end{equation}

with initial condition $\phi_A^{(l, -)}(\textbf{k}) \equiv \phi^{(l, -)}(\textbf{k}, t_A)$ and TEO \cite{Wollensak1}

\begin{equation}
\label{Gl14}
K_{\textbf{k}}^{(-)}(t|t_A)  =  \sum_{n=0}^{\infty} \left( \begin{array}{rr}  I_n(s)   &  0 \  \  \   \\   0 \  \  \   &  I^{\ast}_n(s)   \end{array} \right) 
\left( \begin{array}{rr} 0  \  \  \  &  \frac{k_2 + ik_1}{\kappa}  \\   - \frac{k_2 - ik_1}{\kappa}  &  0 \  \  \  \end{array} \right)^n  ,
\end{equation}

where $I_0(s) \equiv 1$ and

\begin{equation}
\label{Gl15}
I_n(s)  = \left(  \frac{\kappa s^\delta}{\mu} \right)^n  \int\limits_{\sigma_A}^1 d\sigma_1 \int\limits_{\sigma_A}^{\sigma_1} d\sigma_2 ...\int\limits_{\sigma_A}^{\sigma_{n-1}}    d\sigma_n 
\prod\limits_{m = 1}^{n} \sigma_m^{\delta -1} 
e^{i (- 1)^m \sigma_m \tau}
\end{equation}

for $n \geq 2$. The lower limit of integration is given by $\sigma_A(t) := \tau_A/\tau \equiv s(t_A)/s(t)$, and $\kappa \, := \, \sqrt{k_1^2 + k_2^2}$. The form of the general solutions for massless fermion mode functions in pBI spacetimes (\ref{Gl10}), eq.s (\ref{Gl13}) - (\ref{Gl15}), is reminiscent of an earlier result by Tsamis and Woodard, where solutions for massless scalar mode functions in flat FLRW backgrounds with arbitrary expansion factor had been determined \cite{Woodard}. The transfer matrix introduced in that work as a time-ordered product of the exponential of a line integral corresponds to the above TEO, which also can be written as a time-ordered exponential.

An exact computation of (\ref{Gl14}), (\ref{Gl15}) in the cases $\delta = 1$ and $k_3 = 0$ is feasible, an analytical approximate result when $0 <\delta \leq 1/2 \ (\mu > 0)$ and $\delta = 1 - \epsilon \, (0 \leq \epsilon \ll 1)$ has been calculated in \cite{Wollensak1}.
\\
\section*{III. A TRANSFORMATION GENERATING EXACT AND APPROXIMATE SOLUTIONS}
We consider now the action of the following transformation of the line element parameters $\mu, \, \nu$ with real transformation parameter $a \, \neq \, 0$ :

\begin{equation}
\label{Gl16}
\mu \, \rightarrow \, \mu^{ '} = a \mu, \ \ \ \ \ \nu \, \rightarrow \, \nu^{'} = 1 - a (1 - \nu)   ,
\end{equation}

on $ds^2$ given above:

\begin{equation}
\label{Gl17}
ds^2 \rightarrow ds^{' \, 2} = dt^2 - t^{2 \nu^{'}} (dx^1)^2 -  t^{2 \nu^{'}} (dx^2)^2 -  t^{2 - 2 \mu^{'}} (dx^3)^2  ,
\end{equation}

and on the exact (negative chirality) Weyl-spinor solutions $\phi^{(l, -)}$ :

\begin{equation}
\label{Gl18}
\phi^{(l, -)}(\textbf{k}, t) \rightarrow \phi^{' \, (l, -)}(\textbf{k}, t) = \phi^{(l, -)}(\textbf{k}/a, t^a)   ,
\end{equation}

where the latter transformation follows from eq.s (\ref{Gl13}) - (\ref{Gl15}). The crucial property of this transformation consists in that the transformed spinor $\phi^{'}$ is basically given by the original spinor $\phi$ (apart from the scaled wavevektor $\textbf{k}^{'} = \textbf{k}/a$ and the replacement of $t$ by $t^a$), since the parameter transformation (PT) defined by (\ref{Gl16}) leaves $\delta$ unchanged. Hence, if the $\phi^{(l, -)}$ are exact Weyl-spinor solutions of the system (\ref{Gl6}) valid in a background spacetime described by (\ref{Gl10}), then the spinors $\phi^{' (l, -)}$ given by (\ref{Gl18}) denote the corresponding exact solutions of

\begin{equation}
\label{Gl19}
\partial_t \phi^{' (l, -)} -  \Omega^{' (-)} \phi^{' (l, -)} = 0 
\end{equation}

with

\begin{equation}
\label{Gl20}
\Omega^{' (-)}(\textbf{k}, t)  = a t^{a - 1}  \, \Omega^{(-)}(\textbf{k}/a, t^a)   ,
\end{equation}

valid in backgrounds (\ref{Gl17}). For later convenience we define the equivalence classes of pBI spacetimes (\ref{Gl10}) and of related exact Weyl-spinor solutions by

\begin{equation}
\label{Gl22}
\begin{aligned}
\boldsymbol{[\,\delta\,]} \, =& \, \left\lbrace ds^2 | 1 - \nu = \mu \, \delta, \ \mu, \nu \in \mathbb{R}, \, \mu \neq 0 \right\rbrace  ,
\\
\boldsymbol{[\,\bar{\delta}\,]} \, =& \, \left\lbrace   \phi^{(l, \mp)}(\textbf{k}, t) \, | ds^2 \in \boldsymbol{[\,\delta \,]}  \right\rbrace 
\end{aligned}
\end{equation}

 $(|\delta| < \infty)$, where the $\phi^{(l, \mp)}$ are exact solutions of (\ref{Gl6}), (\ref{Gl7}). For $\delta \neq 1$, not any two elements of $\boldsymbol{[\,\delta\,]}$ are Weyl-related, i.e. $ds^{' \, 2}  \neq  \Lambda^2 \, ds^2$ always, whereas in the special case $\delta = 1$ the opposite is true: $\boldsymbol{[\,1\,]}$  consists of all conformally flat line elements. Since $|g| \equiv  t^{4 \nu + 2 (1 - \mu)}$ transforms with PT (\ref{Gl16}) by

\begin{equation}
\label{Gl23}
|g(t)| \, \rightarrow \, |g^{'}(t)| \, = \, t^{6 (1 - a)} \, |g(t)|^a   ,
\end{equation}

$P_3$ by

\begin{equation}
\label{Gl24}
P_3(k_3, t; t_{\widetilde{A}})  \rightarrow    P^{'}_3(k_3, t; t_{\widetilde{A}})  =  P_3\left( \frac{k_3}{a}, t^a; 
t^{a}_{\widetilde{A}} \right)   ,
\end{equation}

and $\phi^{(l, -)}$ as in (\ref{Gl18}), it follows with (\ref{Gl8}), that $\psi_\textbf{k} ^{(l, \mp)}(\textbf{x},t) \, \rightarrow \,  \psi_\textbf{k} ^{' (l, \mp)}(\textbf{x},t)$, where

\begin{equation}
\label{Gl25}
\psi_\textbf{k}^{' (l, \mp)}(\textbf{x},t)  =  c^{' (l, \mp)}_\textbf{k} \,  
\frac{e^{i\textbf{k} \textbf{x}}}{|g^{'}(t)|^{\frac{1}{4}}}  \,\left( \begin{array}{c} 
e^{\pm i P_3^{'}(k_3, t; t_{\widetilde{A}})} \, \phi_1^{' (l, \mp)}(\textbf{k}, t) 
\\
e^{\mp i P_3^{'}(k_3, t; t_{\widetilde{A}})} \, \phi_2^{' (l, \mp)}(\textbf{k}, t)
\\
\mp \, e^{\pm i P_3^{'}(k_3, t; t_{\widetilde{A}})} \, \phi_1^{' (l, \mp)}(\textbf{k}, t)
\\
\mp \,  e^{\mp i P_3^{'}(k_3, t; t_{\widetilde{A}})} \, \phi_2^{' (l, \mp)}(\textbf{k}, t)
\end{array} \right)
\end{equation} 

($l = 1, \, 2$). To summarize: Let $\psi^{(l, -)}$ be an exact massless bispinor solution of Dirac's eq. (\ref{Gl2}) with background spacetime defined by (\ref{Gl10}). Then $\psi^{' (l, \mp)}$ denotes for arbitrary real $a \neq 0$ an exact massless bispinor solution, where the background is given by (\ref{Gl17}). Both spacetimes are elements of the same equivalence class $\boldsymbol{[\,\delta\,]}$.

This result can also be obtained by use of the coordinate transformation

\begin{equation}
\label{Gl26}
f_0 (t) = t^a \equiv \tilde{x}_0, \ \ f_j (x_j) = a \, x_j \equiv \tilde{x}_j 
\end{equation}

($x_0 \equiv t \neq 0$), with additional constraint

\begin{equation}
\label{Gl26a}
\boldsymbol{\tilde{k} \, \tilde{x}} = \boldsymbol{k \, x}   .
\end{equation}

This constrained transformation is consistent with PT (\ref{Gl16}): Firstly, owing to (\ref{Gl26}) and (\ref{Gl26a}), $x_\mu \rightarrow \tilde{x}_\mu$ leads to the following analog of (\ref{Gl18}): $\phi^{(l, -)}(\textbf{k}, t) \rightarrow \tilde{\phi}^{(l, -)}(\textbf{k}, t) = \phi^{(l, -)}(\textbf{k}/a, t^a)$. Secondly, line element (\ref{Gl10}) transforms as

\begin{equation}
\label{Gl27}
ds^2 \, \rightarrow \, d\tilde{s}^2 \, = \, \Lambda^2 \, ds^{' \, 2}
\end{equation}

with $ds^{' \, 2}$ given in eq (\ref{Gl17}). $d\tilde{s}^2$ is $not$ a pBI line element, but $d\tilde{s}^2$ and $ds^{' \, 2}$ are obviously Weyl-related with scale factor $\Lambda(t) = a \, t^{a - 1}$. Finally, (\ref{Gl26}), (\ref{Gl26a}) applied to (\ref{Gl8}) yields:

\begin{equation}
\psi_\textbf{k} ^{(l, \mp)}(\textbf{x},t) \, \rightarrow \,  \tilde{\psi}_\textbf{k} ^{(l, \mp)}(\textbf{x},t) \nonumber
\end{equation}

with
\begin{equation}
\label{Gl28}
\tilde{\psi}_\textbf{k} ^{(l, \mp)}(\textbf{x},t)  =  \tilde{c}^{(l, \mp)}_\textbf{k} \, 
\frac{e^{i\textbf{k} \textbf{x}}}{|g(t)|^{\frac{a}{4}}}  \,
\left( \begin{array}{c} 
e^{\pm i P_3\big(  \frac{k_3}{a}, t^a; t^{a}_{\widetilde{A}} \big)  } \, \phi_1^{(l, \mp)}\left( \frac{\textbf{k}}{a}, t^a \right) 
\\
e^{\mp i P_3\big(  \frac{k_3}{a}, t^a; t^{a}_{\widetilde{A}} \big)} \, \phi_2^{(l, \mp)}\left( \frac{\textbf{k}}{a}, t^a \right)
\\
\mp \, e^{\pm i P_3\big(  \frac{k_3}{a}, t^a; t^{a}_{\widetilde{A}} \big)} \, \phi_1^{(l, \mp)}\left( \frac{\textbf{k}}{a}, t^a \right)
\\
\mp \,  e^{\mp i P_3\big(  \frac{k_3}{a}, t^a; t^{a}_{\widetilde{A}} \big)} \, \phi_2^{(l, \mp)}\left( \frac{\textbf{k}}{a}, t^a \right)
\end{array} \right)  .
\end{equation}  

Comparing (\ref{Gl28}) with (\ref{Gl25}) one finds: $\tilde{\psi}_\textbf{k} ^{(l, \mp)} = \Lambda^{- 3/2} \, \psi_\textbf{k} ^{' (l, \mp)}$ which is the correct behavior of the bispinors w.r.t. Weyl-scaling, implying that eq.s (\ref{Gl27}), (\ref{Gl28}) are compatible with the earlier outcome, eq.s (\ref{Gl17}), (\ref{Gl25}). As a result, the PT is (in the massless case) equivalent to the special diffeomorphism (\ref{Gl26}) together with constraint (\ref{Gl26a}) and appropriate Weyl-scaling.

Recently, a similar looking method has been presented which relates exact scalar field solutions in different flat FLRW backgrounds, provided one can solve a nonlinear second order differential equation \cite{Lochan}. An analytical solution of this DE is possible if one considers massless scalar fields in flat FLRW universes with power-law behavior ($ds^2 \in  \boldsymbol{[\, 1 \,]}$). In this case these fields are related to massive scalar fields in a de Sitter background.

For further study of (\ref{Gl16}) it is convenient to consider the bijective map $u_a: \, U \, \rightarrow \, U \ (U \subset \mathbb{R}^2 , \ a \, \neq \, 0$):

\begin{equation}
\label{Gl35}
u_a : \left( \begin{array}{c} \mu  \\ \nu  \end{array} \right)  \, \rightarrow \,
\left( \begin{array}{c} a \mu  \\ 1 - a (1 - \nu)  \end{array} \right)   ,
\end{equation}

where for the time being $\mu \neq 0$. This map defines a simple abelian $1$-parameter group $G := \{ u_a \}_{a \neq 0}$ with $u_{a_1} \circ u_{a_2} \, = \, u_{a_1 a_2}$, induced by PT (\ref{Gl16}). Every point $(\mu, \nu) \in U$ represents a pBI universe described by spacetime (\ref{Gl10}). Each equivalence class $\boldsymbol{[\,\delta \,]}$ corresponds to an orbit of $G$ and can be visualized in $U$ as a straight line defined by $\nu(\mu) = - \delta \mu + 1$. 

The n-th repeated action of $u_a$ on an arbitrary point $(\mu, \nu) \in U$ yields:

\begin{equation}
\label{Gl36}
u_a \circ ... \circ u_a \left( \begin{array}{c} \mu  \\ \nu  \end{array} \right) \,  =  \,  \left( \begin{array}{c} a^n \mu  \\ 1 - a^n (1 - \nu)  \end{array} \right)   .
\end{equation} 

When n tends to infinity, the r.h.s. of (\ref{Gl36}) approaches for $|a| <1$ the point $(0, 1)$. Thus, $u_0$ can be defined  as the limiting value $u_0 = \lim_{a \to 0} \, u_a$ (but $u_o \notin G)$, where $u_0(\mu, \, \nu) \equiv (0, \, 1)$
denotes the (only) fixed point of $u_a$. All orbits intersect in this point. Interestingly, the line element associated with this fixed point is given by: $ds^2_{\mathrm{FLRW}} \, = \, dt^2 \, - \, t^2 \, (dx^2 + dy^2 + dz^2)$. This line element describes a special conformally flat FLRW universe which is, however, no element of any equivalence class $\boldsymbol{ [\,\delta \,]}$.

The above considerations allow the generalization of (\ref{Gl22}) by including

\begin{equation}
\label{Gl39}
\boldsymbol{ [\,\infty \,]} \, = \, \{ ds^2 | \nu \in \, \mathbb{R}  \setminus  \{1\}, \, \mu = 0  \}   .
\end{equation}

As a consequence, the maximum domain $U$ of $u_a$ is given by $\mathbb{R}^2  \setminus  \{ (0, 1) \}$.
\\
\section*{IV. EXACT SOLUTIONS}
We consider first the conformally flat class $\boldsymbol{[\, 1 \,]}$ and choose as representative of $\boldsymbol{[\, \bar{1} \,]}$ that exact Weyl-spinor solution $\phi^{(l, -)}$ with $\mu = 1/2 = \nu$. The corresponding line element (\ref{Gl10}) describes a radiation dominated universe. By virtue of eq.s (\ref{Gl16}) - (\ref{Gl18}), the general solutions $\in \boldsymbol{[\, \bar{1} \,]}$ can be written as:

\begin{equation}
\label{Gl40}
\phi^{' (l, -)}(\textbf{k}, t) \, = \, \left( \begin{array}{c} e^{- 4 i \frac{k_3}{a} \sqrt{t_A^{a}} } \, 
e^{2 i [(- 1 )^l \, k \, \mathrm{sign} k_3 \, - \, k_3 ][\sqrt{t^{a}} - \sqrt{t_A^{a}} \,]/a } 
\\ i \, \frac{(- 1)^l \, k\, \mathrm{sign} k_3 \, - \, k_3 }{k_2 + i k_1} \, e^{- 4 i \frac{k_3}{a} \sqrt{t^{a}_{\widetilde{A}} } } \, 
e^{2 i [(- 1 )^l \, k \, \mathrm{sign} k_3 \, + \, k_3 ][\sqrt{t^{a}} - \sqrt{t^{a}_A} \,]/a }
\end{array}   \right)
\end{equation}

($k := \sqrt{\kappa^2 + k^2_3} \equiv \sqrt{k_1^2 + k_2^2 + k_3^2}$). For $a = 1$ one recovers the exact result of the case $\mu = \nu = 1/2$ \cite{Wollensak1}. With (\ref{Gl5}), (\ref{Gl24}) and (\ref{Gl40}) holds then

\begin{equation}
\label{Gl41}
\left( \begin{array}{c} 
e^{ i P^{'}_3  } \ \phi_1^{' (l, -)}
\\
e^{- i P^{'}_3 } \ \phi_2^{' (l, -)}
\end{array}   \right)
\, = \, c_{k_3}^{' (l, -)} \, e^{ 2 i \, \frac{(-1)^l k \, \mathrm{sign} k_3}{a} \, [t^{a/2} - t_A^{a/2} \,] } \,
\left( \begin{array}{c} 
1 
\\
i \, \frac{(- 1)^l \, k\, \mathrm{sign} k_3 \, - \, k_3 }{k_2 + i k_1}
\end{array}   \right)   .
\end{equation}

Eq. (\ref{Gl25}) with (\ref{Gl23}) yields:
\begin{equation}
\label{Gl42}
\psi_\textbf{k} ^{' (l, -)}(\textbf{x},t)  =  c^{' (l, -)}_\textbf{k} \, e^{i\textbf{k} \textbf{x}} \,  e^{ 2 i \, \frac{(-1)^l k \, \mathrm{sign} k_3}{a} \, [t^{a/2} - t_A^{a/2} \,] } \,
\frac{t^{- \frac{3}{2} (1 - a)}}{|g(t)|^{\frac{a}{4}}} 
\left( \begin{array}{c} 
1
\\
i \,\frac{(- 1)^l \, k\, \mathrm{sign} k_3 \, - \, k_3 }{k_2 + i k_1}
\\
- \,1
\\
- i \, \frac{(- 1)^l \, k\, \mathrm{sign} k_3 \, - \, k_3 }{k_2 + i k_1}
\end{array} \right)
\end{equation}

($|g| = t^3$), representing for any $a \neq 0$ the exact massless negative chirality bispinor solutions in background spacetimes (\ref{Gl10}) with $\mu = a/2, \, \nu = 1 - a/2$. The positive chirality bispinor solutions can be inferred from (\ref{Gl9}).

In the next example we seek solutions of the Weyl-Dirac-equation when the background spacetime is described by an arbitrary line element $ds^2 \in$ \textbf{[\,1/2\,]}. On choosing $\mu = 2, \, \nu = 0$ one gets with (\ref{Gl10})

\begin{equation}
\label{Gl29}
ds^2 = dt^2 - (dx^1)^2 -  (dx^2)^2 -  t^{- 2} (dx^3)^2  ,
\end{equation}

and with (\ref{Gl6}), (\ref{Gl7}) 

\begin{equation}
\label{Gl30}
\begin{aligned}
\partial_t \phi_1^{(l, -)} \, = \, & (i k_1 + k_2)\, e^{- i k_3 (t^2 - t^2_{\widetilde{A}})} \, \phi_2^{(l, -)}   ,
\\
\partial_t \phi_2^{(l, -)} \, = \, & (i k_1 - k_2)\, e^{ i k_3 (t^2 - t^2_{\widetilde{A}})} \, \phi_1^{(l, -)}   .
\end{aligned}
\end{equation}

The solutions of this system are \cite{Kamke}:

\begin{equation}
\label{Gl31}
\begin{aligned}
\phi_1^{(1, -)}(\textbf{k}, t) \, = \,& z^{- 1/4} \, e^{z/2} \, W_{- b; \frac{1}{4}}(- z)   ,
\\
\phi_2^{(1, -)}(\textbf{k}, t) \, = \,& 2 \, \frac{\sqrt{- i k_3}}{i k_1 + k_2} \, z^{- 3/4} \, e^{- z/2} \, 
\left[ \frac{i \kappa^2}{4 k_3} \, W_{- b; \frac{1}{4}}(- z) \ - \  W_{ 1 - b; \frac{1}{4}}(- z) \right]
\end{aligned}
\end{equation}

(with $z := - i k_3 t^2, \, b := (1 + i \kappa^2/k_3)/4$), and:

\begin{equation}
\label{Gl32}
\begin{aligned}
\phi_1^{(2, -)}(\textbf{k}, t) \, = \,& z^{- 1/4} \, e^{z/2} \, W_{b; \frac{1}{4}}(z)   ,
\\
\phi_2^{(2, -)}(\textbf{k}, t) \, = \,&  i \, \frac{(- i k_1 + k_2) \, \mathrm{sign}k_3}{2 \sqrt{ i k_3}} \, z^{- 3/4} \, e^{- z/2} \, 
\left[ W_{b; \frac{1}{4}}(z) \ + \ \frac{\frac{i \kappa^2}{2 k_3} - 1}{2} \, W_{b - 1; \frac{1}{4}}(z) \right]
\end{aligned}
\end{equation}

($W_{\alpha; \beta}(z)$ denotes Whittaker's function, and for simplicity we have put $t_A = 0$). The two independent negative chirality bispinors are given by (\ref{Gl8}) with $|g| = t^{- 2}$ and $P_3(k_3,t; 0) = k_3 t^2/2$, and the corresponding positive chirality bispinors can be determined with the help of (\ref{Gl9}).

For any spacetime obtained from (\ref{Gl29}) by virtue of (\ref{Gl16}), the exact spinor solutions of (\ref{Gl19}) are according to (\ref{Gl18}) found to be $\phi^{(l, -)}(\textbf{k}/a, t^a)$, where the $\phi^{(l, -)}(\textbf{k}, t)$ have been calculated in eq.s (\ref{Gl31}) and (\ref{Gl32}), resp. The pertaining bispinors follow from (\ref{Gl25}). It is for example straightforward to derive the exact solutions for massless bispinors propagating in a stiff-fluid background spacetime. This spacetime is an exact solution of Einstein`s field equations, when the material content is described by the "perfect fluid" stress tensor $T_{\alpha \beta} = (\varepsilon + p) u_{\alpha} u_{\beta} \, + \, p g_{\alpha \beta}$. $p$ denotes the pressure, $\varepsilon$ the energy density, and $u_{\alpha}$ the four velocity. The equation of state reads $p = (\gamma - 1) \varepsilon$ with $\gamma$ a constant. $\gamma = 2$ is the stiff-matter case where the speed of sound equals the speed of light. In this case the line element is given by (\ref{Gl17}) with $\mu^{'} = 1, \,  \nu^{'} = 1/2$ \cite{Kramer}:
\begin{equation}
\label{Gl33}
ds^{' \,2} = dt^2 - t (dx^1)^2 - t (dx^2)^2 -  (dx^3)^2  .
\end{equation}

Clearly, $ds^{' \, 2}$ belongs to the class \textbf{[\,1/2\,]}, too. On comparing (\ref{Gl29}), (\ref{Gl33}), one immediately finds with (\ref{Gl16}): $a = 1/2$. By virtue of $|g'| = t^2$, $P_3(2 k_3,\sqrt{t};0) = k_3 t$ and (\ref{Gl31}), (\ref{Gl32}) one gets from (\ref{Gl25}) the two solutions:

\begin{equation}
\label{Gl34}
\psi_\textbf{k} ^{' (l, -)}(\textbf{x},t)  =  c^{' \, (l, -)}_\textbf{k} \, \frac{e^{i\textbf{k} \textbf{x}}}{\sqrt{t}} \,\left( \begin{array}{c} 
e^{ i k_3 t  } \ \phi_1^{(l, -)}\left( 2 \textbf{k}, \sqrt{t} \right) 
\\
e^{- i k_3 t} \ \phi_2^{(l, -)}\left( 2 \textbf{k}, \sqrt{t} \right)
\\
- \, e^{ i k_3 t} \ \phi_1^{(l, -)}\left( 2 \textbf{k}, \sqrt{t} \right)
\\
- \,  e^{- i k_3 t} \ \phi_2^{(l, -)}\left( 2 \textbf{k}, \sqrt{t} \right)
\end{array} \right)  .
\end{equation}

These are the exact massless negative chirality bispinor mode solutions in a stiff-fluid background with line element (\ref{Gl33}) \cite{Wollensak1}. Again, the positive chirality solutions are obtained utilizing (\ref{Gl9}). The general bispinors are given by (\ref{Gl25}) together with (\ref{Gl44}), and the general line element reads:

\begin{equation}
ds^{2} = dt^2 - t^{2 - a} [(dx^1)^2 +  (dx^2)^2] -  t^{2 - 2 a} (dx^3)^2   .
\nonumber
\end{equation}

There are further exactly solvable equivalence classes: For the class $\boldsymbol{[\,\overline{\infty} \,]}$ one gets the general negative chirality Weyl-spinor solutions:

\begin{equation}
\label{Gl52a}
\phi^{' (1, -)}(\textbf{k}, t) \,  =  \, \left( \begin{array}{c} 
t^{i k_3}_{\widetilde{A}} \, t^{\frac{a}{2} - i k_3} \, J_{\frac{1}{2} - i \frac{k_3}{a}}\left(\frac{\kappa}{a} \, t^{a} \right)
\\
\frac{\kappa\, t^{- i k_3}_{\widetilde{A}}}{k_2 + i k_1} \, t^{\frac{a}{2} + i k_3} \, J_{- \frac{1}{2} - i \frac{k_3}{a}}\left(\frac{\kappa}{a} \, t^{a} \right)
\end{array}   \right)  ,\ \
\phi^{' (2, -)} \, = \,  \left( \begin{array}{c}  \left( \phi_2^{' (1, -)} \right)^\ast    \\ -  \, \left( \phi_1^{' (1, -)} \right)^\ast  \end{array} \right)   \nonumber
\end{equation}

($J_\nu$ denotes Bessel's function). These solutions solve

\begin{equation}
\label{Gl52b}
\begin{aligned}
\partial_t \phi_1^{' (l, -)} \, = \, &  \frac{i k_1 + k_2}{t^{1 - a}} \, \left( \frac{t}{ t_{\widetilde{A}}} \right)^{- 2 i k_3}  \, \phi_2^{' (l, -)}   ,
\\
\partial_t \phi_2^{' (l, -)} \, = \, & \frac{i k_1 - k_2}{t^{1 - a}} \, \left( \frac{t}{ t_{\widetilde{A}}} \right)^{+ 2 i k_3} \, \phi_1^{' (l, -)}   .
\end{aligned}   \nonumber
\end{equation}

The pertaining general expression for the line elements of $\boldsymbol{[\,\infty \,]}$ reads: $ds^{' 2} = dt^2 - t^{2 - 2 a} (dx^2 +  dy^2) -  t^2 dz^2$. Finally, the four bispinor solutions are given by (\ref{Gl25}), (\ref{Gl9}) and (\ref{Gl23}) (we chose $\mu = 0 = \nu$, hence $|g| = t^2$). The special case $a = 1$  has been studied in ref. \cite{Shishkin}. The corresponding spacetime represents the planar (flat) Kasner solution with Kasner exponents $(0, 0, 1)$.

For the class $\boldsymbol{[\,\bar{0} \,]}$ one obtains the solutions:

\begin{equation}
\label{Gl52d}
\phi^{' (1, -)}(\textbf{k}, t) \,  =  \, t^{- \frac{a}{2}} \left( \begin{array}{c} 
e^{- \frac{z_a}{2}} \, W_{- \frac{1}{2}; i \, \frac{\kappa}{a}}(z_a)
\\
- \frac{a \, e^{- (z_a)_{\widetilde{A}}  } }{k_2 + i k_1} \,
e^{\frac{z_a}{2}} \, W_{\frac{1}{2}; i \, \frac{\kappa}{a}}(z_a)
\end{array}   \right),\ \ \
\phi^{' (2, -)} \, = \,  \left( \begin{array}{c}  \left( \phi_2^{' (1, -)} \right)^\ast    \\ -  \, \left( \phi_1^{' (1, -)} \right)^\ast  \end{array} \right)   \nonumber
\end{equation}

($z_a \equiv  + 2 i k_3 t^a/a, \,   (z_a)_{\widetilde{A}} \equiv  + 2 i k_3 t_{\widetilde{A}}^a/a$), which solve:

\begin{equation}
\label{Gl52e}
\begin{aligned}
\partial_t \phi_1^{' (l, -)} \, = \, &  \frac{i k_1 + k_2}{t} \  e^{- 2 i \frac{k_3}{a} \, (t^a - t_{\widetilde{A}}^a)}  \, \phi_2^{' (l, -)}   ,
\\
\partial_t \phi_2^{' (l, -)} \, = \, & \frac{i k_1 - k_2}{t} \ e^{2 i \frac{k_3}{a} \, (t^a - t_{\widetilde{A}}^a)} \, \phi_1^{' (l, -)}   .
\end{aligned}   \nonumber
\end{equation}

The general line element in $\boldsymbol{[\,0 \,]}$ reads: $ds^{' 2} = dt^2 - t^2 (dx^2 +  dy^2) -  t^{2 - 2 a} dz^2$, 
and the four bispinor solutions are again found with the help of eq.s (\ref{Gl25}), (\ref{Gl9}), (\ref{Gl23}) (we chose $\mu = 1 = \nu$, hence $|g| = t^4$). The special case $a = 2$  has been treated in \cite{Pimentel}.

The above given complete sets of exact mode solutions $\psi^{'}_{\textbf{k}}$ can in principle be used to construct fermion propagators in the respective spacetimes and then to calculate the one loop effective action, similar to earlier work \cite{Candelas2}. However, while the massless fermion (Feynman) propagator in flat FLRW spacetimes can be easily computed owing to commuting tangent vectors \cite{Koksma}, the analogue calculation in any pBI spacetime $\in [\boldsymbol{\delta}] \setminus [\boldsymbol{1}]$ will be more complicated. The reason is that the $e_\mu$ appearing in the covariant differential of eq. (\ref{Gl2}) are now noncommuting vector$fields$ satisfying $[e_\mu, e_\nu] = C^\lambda_{\mu \nu} \, e_\lambda$. 

It has been stated earlier that the PT maps every line element (\ref{Gl10}) of any class $\boldsymbol{[\,\delta \,]}$ for vanishing $a$ into $ds^2_{\mathrm{FLRW}}$. This must also be true for the corresponding exact bispinor solutions, i.e. $\lim_{a \to 0} \,\psi^{'} =  \psi_{\mathrm{FLRW}}$, and will be verified in the appendix for the exactly soluble cases $\delta = 1, \, 1/2$. Remarkably, it can be shown that for the approximate TEO solutions of ref. \cite{Wollensak1} also holds: $\lim_{a \to 0} \,\psi^{'}_{\mathrm{TEO}} = \psi_{\mathrm{FLRW}}$. 
\\
\section*{V. APPROXIMATE SOLUTIONS}
Exact solutions are in many cases not at disposal. However, the method developed in sect. III can also be used to obtain approximate solutions of a given model with background spacetime in $\boldsymbol{[\, \delta \,]}$. As before in the exact case, one considers first a (possibly) simpler model with background belonging to the same class $\boldsymbol{[\, \delta \,]}$. Having then derived the approximate solutions of this model, one gets by use of PT (\ref{Gl16}) in complete analogy to the treatment of the exact case in sect. IV. the desired approximate solutions of the initial model and in fact of the entire class $\boldsymbol{[\, \overline{\delta} \,]}$.

As an illustrative example we want to determine the approximate bispinor solutions of the Weyl-Dirac-equation with axisymmetrical Kasner background ($\mu = 4/3, \, \nu = 2/3$, i.e. $\delta = 1/4$):
\begin{equation}
\label{Gl56}
ds^2 = dt^2 - t^{4/3} (dx^1)^2 -  t^{4/3} (dx^2)^2 -  t^{- 2/3} (dx^3)^2  .
\end{equation}

This vacuum Kasner solution evolves for $t \rightarrow 0$ as the preinflationary limiting case of a general BI-background geometry with axial symmetry \cite{Gm}.

From (\ref{Gl6}), (\ref{Gl7}) together with 

\begin{equation}
\label{Gl54}
\phi_1^{(l, -)}(\textbf{k}, t) \,  = \, t^{- \nu/2} \, e^{- i (\tau - \tau_{\widetilde{A}})/2} \, \varphi(\textbf{k}, t)   ,
\end{equation}

follows the second order ordinary differential equation (ODE):

\begin{equation}
\label{Gl55}
\partial^2_t \, \varphi + \left[ - \frac{(\nu + 2 i k_3 t^\mu)^2}{4 t^2} \, + \, \frac{\nu - 2 i k_3 (\mu - 1) \, t^\mu}{2 t^2} \, + \, \frac{\kappa^2}{t^{2 \nu}}
\right] \, \varphi \, = \, 0   .
\end{equation}

For $\delta \neq 0$ one can always put $\nu = 0$, because each equivalence class $\boldsymbol{[\, \delta\,]}$ possesses exactly one such representative $(\mu = 1/\delta, \, \nu = 0)$. 

Spacetime (\ref{Gl56}) is an element of the class $\textbf{[ 1/4 ]}$. In the first step we therefore choose as representative the line element (\ref{Gl10}) with $\mu = 4, \, \nu = 0$:

\begin{equation}
\label{Gl57}
ds^2 = dt^2 -  (dx^1)^2 -  (dx^2)^2 -  t^{- 6} (dx^3)^2  .
\end{equation}

Insertion of

\begin{equation}
\label{Gl58}
\varphi(\textbf{k}, t) \, = \, C \, \exp \left( \int_{t_{\widetilde{A}}}^{t} u(\textbf{k}, y) dy  \right)
\end{equation}

into (\ref{Gl55}) yields the Riccati-type nonlinear DE:

\begin{equation}
\label{Gl59}
\begin{aligned}
&\partial_t u + u^2 - \mathcal{K} = 0   ,
\\
&\mathcal{K}(\textbf{k}, t) = 3 i k_3 \, t^2 - (k_3^2 \, t^6 + \kappa^2)   .
\end{aligned}
\end{equation}

It can be asymptotically solved with ansatz

\begin{equation}
\label{Gl60}
u_0(\textbf{k}, t) = \sqrt{\mathcal{K}(\textbf{k}, t)}   .
\end{equation}

Here and in the following the square root is understood to be the principal branch. This solution can be improved by the iteration $u_{n + 1} = \sqrt{\mathcal{K} - \partial_t u_n}$, ($n \geq 0$)  \cite{Berg}. For example, the first and second iteration are given by:
\begin{equation}
\label{Gl62}
\begin{aligned}
u_1(\textbf{k}, t) \, =& \, i \, \mathrm{sign} k_3 \, \sqrt{k_3^2 t^6 + \kappa^2 - \frac{3}{2 \, t^2} - i \, \frac{3 \kappa^2}{2 k_3 t^4} + O\left( \frac{1}{k_3 t^6}  \right)}   ,
\\
u_2(\textbf{k}, t) \, = & \, i \, \mathrm{sign} k_3 \, \sqrt{k_3^2 t^6 + \kappa^2 - i \, \frac{3 \kappa^2}{2 k_3 \, t^4} + i \, \frac{15}{4 k_3 t^6} + O\left( \frac{\kappa^2}{k_3^2 t^8}  \right)}   .
\end{aligned}
\end{equation}

With (\ref{Gl58}) substituted into (\ref{Gl54}) one finds:

\begin{equation}
\label{Gl63}
\phi_1^{(2, -)}(\textbf{k}, t) \,  =  \, e^{- i \frac{k_3}{4} t^4} \, \exp \left(  i \, \mathrm{sign} k_3 \, \int_{t_{\widetilde{A}}}^{t} dy \, \sqrt{k_3^2 y^6 + \kappa^2 + \frac{\mathcal{A}_n}{ y^\ell} } \,  \right)   ,
\end{equation}

where $\mathcal{A}_n$ and $\ell(n)$ depend on the level of iteration, e.g.:
\begin{equation}
n = 1:  \, \ell(1) = 2, \, \mathcal{A}_1 = - \frac{3}{2}; \ \  n = 2:  \, \ell(2)= 4, \, \mathcal{A}_2= - i \, \frac{3 \kappa^2}{2 k_3}   .  \nonumber
\end{equation}

The asymptotic expansion of the integral in (\ref{Gl63}) yields (for $n = 2$):

\begin{equation}
\label{Gl64}
\int_{t_{\widetilde{A}}}^{t}  dy \, \sqrt{k_3^2 y^6 + \kappa^2 + \frac{\mathcal{A}_2}{ y^{\ell(2)}} } \, \, \sim \,  \frac{1}{4} \, |k_3| t^4 \, 
+ \, C_{\textbf{k}}(t_{\widetilde{A}})  \, - \, \frac{\kappa^2}{4 |k_3|} t^{- 2} + ...  .   \nonumber
\end{equation}

With $\phi_2^{(2, -)}  \equiv  \partial_t \, \phi^{(2, -)}_1/\mathcal{P}^{(-)}$, one gets:

\begin{equation}
\label{Gl66}
\phi^{(2, -)}(\textbf{k}, t) \, = \,   \left( \begin{array}{c} 1  \\ 
i \   \frac{\mathrm{sign} k_3 \,\sqrt{k_3^2 t^6 + \kappa^2 + \frac{\mathcal{A}_2}{ t^4} } \  - \  k_3 t^3}{(i k_1 + k_2) \, e^{- i k_3 t^4/2}}  \end{array} \right)  \,  \phi^{(2, -)}_1(\textbf{k}, t)  .
\end{equation}

As before, the other (orthogonal) Weyl-spinor solution is:

\begin{equation}
\label{Gl67}
\phi^{(1, -)}(\textbf{k}, t) \, = \,  \left( \begin{array}{c}  \left( \phi_2^{(2, -)}(\textbf{k}, t) \right)^\ast    \\ -  \, \left( \phi_1^{(2, -)}(\textbf{k}, t) \right)^\ast  \end{array} \right)   .
\end{equation}

The asymptotic bispinor solutions are given by (\ref{Gl8}) with $|g| = t^{- 6}$. 

For $t \ll (\kappa/|k_3|)^{1/3}$, the early-time solutions can be determined by solving the reduced ODE of eq. (\ref{Gl55}): $\partial^2_t \varphi + (\kappa^2 - 3 i k_3 t^2) \varphi = 0$. One gets:
\begin{equation}
\begin{aligned}
\label{Gl72}
\phi^{(1, -)}(\textbf{k},t) \, &= \,  \frac{\sqrt{\beta}}{\xi^{1/4}} \,  \left( \begin{array}{c} W_{\alpha; \frac{1}{4}}(\xi)
\\
- \, \frac{2}{t \, \beta} \left\lbrace  \left[\frac{\xi^2}{6} - \frac{\xi}{2} + \alpha + \frac{1}{4} \right] W_{\alpha;\frac{1}{4}}(\xi) + W_{\alpha + 1;\frac{1}{4}}(\xi)   \right\rbrace
\end{array} \right)
\\
\phi^{(2, -)}(\textbf{k},t) \, &= \,  \left( \begin{array}{c}  \left( \phi_2^{(1, -)}(\textbf{k}, t) \right)^\ast    \\ -  \, \left( \phi_1^{(1, -)}(\textbf{k}, t) \right)^\ast  \end{array} \right)   ,
\end{aligned}
\end{equation}

where
\begin{equation}
\label{Gl73}
\alpha \, = \,\frac{\kappa^2}{4 \sqrt{3 |k_3|}} \, e^{-\,  i \, \frac{\pi}{4} \,  \mathrm{sign}k_3} , \ \ \
\beta(t) \, = \,\frac{k_2 + i k_1}{e^{\xi^2/6}} , \ \ \   \xi(t) \, = \ \sqrt{3 |k_3|} \, e^{i \, \frac{\pi}{4} \,  \mathrm{sign}k_3}  \, t^{2}  .
\nonumber
\end{equation}

The choice of the $\phi^{(l, -)}$  in (\ref{Gl72}) and equally in eq.s (\ref{Gl66}), (\ref{Gl67}) has been made in such a way that it is compatible with the associated TEO result, see the discussion in sect. VI.

Eq.s (\ref{Gl66}) - (\ref{Gl72}) are the approximate solutions of (\ref{Gl6}) ($\mu = 4, \, \nu = 0$):

\begin{equation}
\label{Gl68}
\begin{aligned}
\partial_t \phi_1^{(l, -)}(\textbf{k}, t) \, = \, & (i k_1 + k_2)\, e^{- i \frac{k_3}{2} t^4} \, \phi_2^{(l, -)}(\textbf{k}, t)   ,
\\
\partial_t \phi_2^{(l, -)}(\textbf{k}, t) \, = \, & (i k_1 - k_2)\, e^{ i \frac{k_3}{2} t^4} \, \phi_1^{(l, -)}(\textbf{k}, t)   .
\end{aligned}
\end{equation}

In the second step we turn to the computation of the approximate bispinor solutions with background (\ref{Gl56}). The transition from (\ref{Gl57}) to (\ref{Gl56}) is managed by (\ref{Gl16}) with $a = 1/3$. With (\ref{Gl18}), (\ref{Gl67}), (\ref{Gl66}) the large-time solutions read:

\begin{equation}
\label{Gl74}
\begin{aligned}
\phi^{'  (1, -)}(\textbf{k}, t) \, = &\,   \left( \begin{array}{c}  \left( \phi_2^{'  (2, -)}(\textbf{k}, t) \right)^\ast    \\ -  \, \left( \phi_1^{'  (2, -)}(\textbf{k}, t) \right)^\ast  \end{array} \right)   ,
\\
\phi^{'  (2, -)}(\textbf{k}, t) \, = &\, \left( \begin{array}{c} 1  \\ 
i \  \frac{\mathrm{sign} k_3 \,\sqrt{k_3^2 t^2 + \kappa^2 - \frac{ i  \kappa^2}{2 k_3 t^{4/3}} } \  - \  k_3 t}{(i k_1 + k_2) \, e^{- i \frac{3}{2} k_3 t^{4/3}}}  \end{array} \right)  \,  \phi^{(2, -)}_1(3 \textbf{k}, t^{1/3})  ,
\end{aligned}
\end{equation}

and with (\ref{Gl72}), (\ref{Gl18}) the early-time solutions are

\begin{equation}
\begin{aligned}
\label{Gl75}
\phi^{' (1, -)}(\textbf{k},t) \, &= \,  \frac{\sqrt{\beta^{'}}}{\xi^{' 1/4}} \,  \left( \begin{array}{c} W_{\alpha^{'}; \frac{1}{4}}(\xi^{'})
\\
\frac{- \, 2}{t^{1/3} \, \beta^{'}} \left\lbrace  \left[\frac{\xi^{' 2}}{6} - \frac{\xi^{'}}{2} + \alpha^{'} + \frac{1}{4} \right] W_{\alpha^{'};\frac{1}{4}}(\xi) + W_{\alpha^{'} + 1;\frac{1}{4}}(\xi^{'})   \right\rbrace
\end{array} \right)
\\
\phi^{' (2, -)}(\textbf{k},t) \, &= \,  \left( \begin{array}{c}  \left( \phi_2^{' (1, -)}(\textbf{k}, t) \right)^\ast    \\ -  \, \left( \phi_1^{' (1, -)}(\textbf{k}, t) \right)^\ast  \end{array} \right)   ,
\end{aligned}
\end{equation}
where
\begin{equation}
\alpha^{'} \, = \,\frac{3 \kappa^2}{4 \sqrt{|k_3|}} \, e^{-\,  i \, \frac{\pi}{4} \,  \mathrm{sign}k_3} , \
\beta^{'}(t) \, = \, \frac{3 (k_2 + i k_1)}{e^{\xi^{' 2}/6}}, \   \xi^{'}(t) \, = \ 3 \sqrt{|k_3|} \, e^{i \, \frac{\pi}{4} \,  \mathrm{sign}k_3}  \, t^{2/3}  .
\nonumber
\end{equation}

With (\ref{Gl9}), (\ref{Gl25}) ($|g^{'}| = t^2$) one finds again the corresponding approximate bispinor solutions. Eq.s (\ref{Gl74}), (\ref{Gl75}) represent the approximate solutions of

\begin{equation}
\label{Gl76}
\begin{aligned}
\partial_t \phi_1^{' (l, -)}(\textbf{k}, t) &= \frac{i k_1 + k_2}{t^{2/3}} \, e^{- \frac{3 i}{2}  k_3 t^{4/3}} \, \phi_2^{' (l, -)}(\textbf{k}, t)
\\
\partial_t \phi_2^{' (l, -)}(\textbf{k}, t) &=  \frac{ i k_1 - k_2}{t^{2/3}} \, e^{ \frac{3 i}{2}  k_3 t^{4/3}} \, \phi_1^{' (l, -)}(\textbf{k}, t)   .
\end{aligned} 
\end{equation}

This system can be obtained either from eq.s (\ref{Gl5}) - (\ref{Gl7}) with $\alpha_1 = t^{2/3} = \alpha_2, \ \alpha_3 = t^{- 1/3}$, or from (\ref{Gl68}) by virtue of (\ref{Gl19}), (\ref{Gl20}) with $a = 1/3$. 

It is instructive to compare the PT solutions (\ref{Gl74}), (\ref{Gl75}) with the outcome of the TEO approach. From (\ref{Gl75}) follows for $t \rightarrow 0$:

\begin{equation}
\label{Gl77}
\begin{aligned}
\phi^{' (1, -)}(\textbf{k},t) \, = \ & A_1^{'} (\textbf{k})  \left( \begin{array}{c}      - \frac{3^{3/4}}{2^{5/4}}   \, \frac{\Gamma(\frac{1}{4} - \alpha^{'})}{\Gamma(\frac{3}{4} - \alpha^{'})} \, (k_2 + i k_1) \, \left( \frac{3 i}{2} k_3  \right)^{- 1/4} \, [1 + O(\tau^{1/4})]   \\    1 \, + \, O(\eta_\delta \, \tau^{1/4})  \end{array} \right)
\\
\phi^{' (2, -)}(\textbf{k},t) \, = \ & A_2^{'} (\textbf{k}) \left( \begin{array}{c}  \frac{2^{5/4}}{3^{3/4}} \,  \frac{\Gamma(\frac{3}{4} - \alpha^{' \ast } ) }{\Gamma(\frac{1}{4} - \alpha^{' \ast} )} \, \frac{1}{k_2 - i k_1}   \left[  \left( \frac{3 i}{2} \, k_3  \right)^\frac{1}{4}  \right]^\ast \, [1 \, + \, O(\eta_\delta \, \tau^{1/4}) ] \\  1 \, + \, O(\tau^{1/4})  \end{array} \right)
\end{aligned}
\end{equation}

with
\begin{equation}
\eta_\delta := \frac{\kappa^2}{2 |k_3|^{2 \delta}} 
\end{equation}

$(\delta =  1/4$ from now on) and $\tau$ defined by (\ref{Gl12}), whereas (\ref{Gl74}) assumes at late times the form:

\begin{equation}
\label{Gl78}
\begin{aligned}
\phi_1^{' (1, -)}(\textbf{k},t)  \, \sim  &  \, B_1(\textbf{k}) \, e^{- i \tau} \tau^{- 3/4}   \, [1 + O(\eta_\delta \, \tau^{- 1/2} ) ]
\\
\phi_2^{' (1, -)}(\textbf{k},t)   \, \sim  &  \, 1  +  O(\eta_\delta \, \tau^{- 1/2} )
\\
\phi_1^{' (2, -)}(\textbf{k},t)  \, \sim  &  \,  1  - O(\eta_\delta \, \tau^{- 1/2} )
\\
\phi_2^{' (2, -)}(\textbf{k},t)  \, \sim  &  \, B_2(\textbf{k}) \, e^{i \tau}  \tau^{- 3/4} \, [1 + O(\ \eta_\delta \, \tau^{- 1/2} )]  .
\end{aligned}
\end{equation}

For $t \rightarrow \infty$ holds:

\begin{equation}
\label{Gl79}
\phi^{' (1, -)}(\textbf{k},t) \, \rightarrow  \, \left( \begin{array}{c} 0  \\  1  \end{array} \right)  , \ \   \phi^{' (2, -)}(\textbf{k},t) \, \rightarrow \,   \left( \begin{array}{c} 1  \\  0  \end{array} \right)   .
\end{equation}

To compare (\ref{Gl74}), (\ref{Gl75}) with the solutions of the TEO formalism one must impose (\ref{Gl79}) on the TEO solutions. These asymptotic conditions fix owing to (\ref{Gl13}) in a unique way the TEO solutions. For example, the first component of the asymptotic TEO solution $\phi_{{\mathrm{TEO}}}^{(1, -)}$ reads ($t_A = 0$) \cite{Wollensak1}:

\begin{equation}
\label{Gl80}
\begin{aligned}
\phi_{{\mathrm{TEO}}_1}^{(1, -)}(\textbf{k},t) \, & \equiv \, \left( K_{\textbf{k}}^{(-)}(t|0) \right)_{11}  \phi_{{\mathrm{TEO}}_1}^{(1, -)}(\textbf{k},0) \,  + \, \left( K_{\textbf{k}}^{(-)}(t|0) \right)_{12}  \phi_{{\mathrm{TEO}}_2}^{(1, -)}(\textbf{k},0)
\\
& \sim \,  \Bigg( 1 \, + \, 2 i d \,  \mathcal{D} \,  G(\tau) \,  \Bigg[\  \frac{e^d}{\delta}  \left\lbrace  1 - i \mathcal{D}^2 G(\tau)  \right\rbrace  \,  F_\delta(- d) 
\\
& + \,  \frac{e^{3 d}}{\delta} \, i \mathcal{D}^2  G(\tau) \,  F_\delta(- 3 d)  - \, \frac{e^{- i \tau [1 + \mathcal{D}^2 E_1(\tau)]  }}{ |\tau|^\delta}  \,  \frac{\Gamma(\delta)}{(- i \, \mathrm{sign} k_3)^\delta}\, 
\\
& +  \, i \,  \frac{e^{i \mathcal{D}^2 [E_2(\tau) - E_1(\tau)] }}{\tau [1 + \mathcal{D}^2 E_2(\tau)]}  \Bigg] + ...
\Bigg) \,
\phi_{{\mathrm{TEO}}_1}^{(1, -)}(\textbf{k},0)
\\
& + \, \frac{k_+}{\kappa}  \frac{\sqrt{2 \eta_\delta}}{\mu } \left( \frac{\mu}{2} \right)^\delta \ \Bigg( \  e^{i \mathcal{D}^2 \, \tau E_1(\tau) }    \Bigg[   \frac{\Gamma(\delta)}{(i \, \mathrm{sign} k_3)^\delta}
\\
& + \, i \, \mathrm{sign} k_3 \, \frac{e^{- i \tau [1 + \mathcal{D}^2 \, E_2(\tau)  ]}}{|\tau|^{1 - \delta} \,  [1 + \mathcal{D}^2 \, E_2(\tau)]} \, \Bigg] + ...\Bigg) \, \phi_{{\mathrm{TEO}}_2}^{(1, -)}(\textbf{k},0)
\end{aligned}
\end{equation}

where $F_\delta(x) :=$   $_1F_1(\delta; \delta + 1; x)$ denotes Kummer's function, $d := 1 - \delta, \mathcal{D} := (\delta^{- 1} - 1) F_\delta(- d)/\mathrm{sinh}(d)$, and $\tau \equiv 3 k_3 t^{4/3}/2$ . Furthermore,

\begin{equation}
E_1(\tau) := \, \frac{e^{2 d} - 1}{4 d} \,  \left( \frac{\mu}{2} \right)^{- 2 d } \, 
\frac{\eta_\delta \, |\tau|^{- 2 d }}{1 - i \frac{d}{\tau}}  ,
\  \ 
E_2(\tau) :=  \, \frac{- 2 d E_1(\tau)}{e ^{- 2 d} - 1}  ,
\  \ 
G(\tau) := \frac{\tau E_1(\tau)}{e^{2 d} - 1}     .   \nonumber
\end{equation}

Since (\ref{Gl80}) must satisfy the first asymptotic condition in (\ref{Gl79}), one finds the $initial$ condition:

\begin{equation}
\label{Gl82}
\phi_{\mathrm{TEO}}^{(1, -)}(\textbf{k},0) \, = \  A_1(\textbf{k}) \,  \left( \begin{array}{c}      - \frac{3}{4}   \, \Gamma(\frac{1}{4} )  \, (k_2 + i k_1) \, \left( \frac{3 i}{2} k_3  \right)^{- 1/4}    \\    1    \end{array} \right)   .
\end{equation}

Analogously one gets from the second asymptotic condition in (\ref{Gl79})

\begin{equation}
\label{Gl83}
\phi_{\mathrm{TEO}}^{(2, -)}(\textbf{k},0) \, = \  A_2(\textbf{k}) \, \left( \begin{array}{c}  \frac{4/3 }{\Gamma(\frac{1}{4} )} \, \frac{1}{k_2 - i k_1}   \left[  \left( \frac{3 i}{2} \, k_3  \right)^\frac{1}{4}  \right]^\ast \\  1  \end{array} \right)   .
\end{equation}

Use of the small-$t$-expansion of the TEO defined in eq. (\ref{Gl14}),

\begin{equation}
K_{\textbf{k}}^{(-)}(t| 0) \, = \,
\left( \begin{array}{rr} 1 \ \ \   &  \frac{k_2 + i k_1}{\mu \, \delta} \, s^{\delta}  \\ -  \frac{k_2 - i k_1}{\mu \, \delta} \, s^{\delta}   &  1 \ \ \   \end{array} \right) \, [1 + O(s^{2 \delta})]   \nonumber
\end{equation}

immediately leads to:

\begin{equation}
\label{Gl84}
\begin{aligned}
\phi_{\mathrm{TEO}}^{(1, -)}(\textbf{k},t) \, = &\, A_1(\textbf{k})  \left( \begin{array}{c}      - \frac{3}{4}   \, \Gamma(\frac{1}{4} ) \, (k_2 + i k_1) \, \left( \frac{3 i}{2} k_3  \right)^{- 1/4} \, [1 + O(\tau^{1/4})]   \\    1 \, + \, O(\eta_\delta \, \tau^{1/4})  \end{array} \right)   ,
\\
\phi_{\mathrm{TEO}}^{(2, -)}(\textbf{k},t) \, = &\, A_2(\textbf{k}) \left( \begin{array}{c}  \frac{4/3 }{\Gamma(\frac{1}{4} )} \, \frac{1}{k_2 - i k_1}   \left[  \left( \frac{3 i}{2} \, k_3  \right)^\frac{1}{4}  \right]^\ast \, [1 \, + \, O(\eta_\delta \, \tau^{1/4}) ] \\  1 \, + \, O(\tau^{1/4})  \end{array} \right)   ,
\end{aligned}
\end{equation}

representing the small-$t$-expansions of the two TEO solutions $\phi_{\mathrm{TEO}}^{(l, -)}$. From inspection of eq.s (\ref{Gl77}) and (\ref{Gl84}) it can be easily verified that these TEO solutions differ by a constant term of the form $\mathcal{B}^{(l, -)} + O(\eta_\delta), \, \mathcal{B}^{(l, -)} \approx 1$ in their first components from the $\eta_\delta$-expanded small-$t$ PT solutions $\phi^{' (l, -)}$ given by (\ref{Gl77}). For instance, the prefactor of $\phi_1^{' (1, -)}$ can be written as:

\begin{equation}
\label{84a}
\frac{3^{3/4}}{2^{5/4}}   \, \frac{\Gamma(\frac{1}{4} - \alpha^{'})}{\Gamma(\frac{3}{4} - \alpha^{'})} \, = \, \frac{3}{4}
\, \Gamma\left( \frac{1}{4} \right) \, [\mathcal{B}^{(1, -)} + O(\eta_\delta)]
\end{equation}

($\mathcal{B}^{(1, -)} \equiv (8/3)^{1/4}/\Gamma(3/4) = 1.04281 ..., \, \alpha^{'} = O(\eta_\delta)$), which (almost) agrees to lowest order in $\eta_\delta$ with the corresponding prefactor of the TEO solution $\phi_{{\mathrm{TEO}}_1}^{(1, -)}$. It has been pointed out in ref. \cite{Wollensak1} that in the asymptotic regime holds: $\eta_\delta  \ll 1$, and that the asymptotic TEO calculated there is correct only to lowest order in $\eta_\delta$ (which, however, does not affect the time dependence of the most dominant terms of the TEO in the small- and large-time expansions). This is manifest in eq. (\ref{84a}), where the l.h.s. denotes the PT result, and the r.h.s. represents to lowest order in $\eta_\delta$ the TEO result.

The asymptotic TEO solutions can be obtained as follows: Eq. (\ref{Gl82}) substituted into (\ref{Gl80}) yields for $|\tau| \gg 1$:

\begin{equation}
\label{Gl85}
\phi_{{\mathrm{TEO}}_1}^{(1, -)}(\textbf{k},t)   \, \sim  \,
A^{(1,-)}_1(\textbf{k}) \, \frac{e^{- i \tau}}{|\tau|^{1 - \delta}}  \,  [1 \, + \, O(\eta_\delta \, \tau^{2 \delta - 1}) ] \, [ 1 + O(\eta_\delta)] \, \phi_{{\mathrm{TEO}}_2}^{(1, -)}(\textbf{k},0) 
\end{equation}

and analogously the second component of $\phi_{\mathrm{TEO}}^{(1, -)}$ is given by:

\begin{equation}
\label{Gl86}
\begin{aligned}
\phi_{{\mathrm{TEO}}_2}^{(1, -)}(\textbf{k},t) \, &\equiv \, \left(-  K_{\textbf{k}}^{(-)}(t|0) \right)^
\ast_{12}  \phi_{{\mathrm{TEO}}_1}^{(1, -)}(\textbf{k},0) \,  + \, \left( K_{\textbf{k}}^{(-)}(t|0) \right)^\ast_{11}  \phi_{{\mathrm{TEO}}_2}^{(1, -)}(\textbf{k},0)
\\
&\sim \, [  1  +   O( \eta_\delta \, \tau^{2 \delta - 1}) ]   \, [ 1 \, + \,  O(\eta_\delta) ]  \, \phi_{{\mathrm{TEO}}_2}^{(1, -)}(\textbf{k},0)   ,
\end{aligned}
\end{equation}

where the spinor $\phi_{{\mathrm{TEO}}_2}^{(1, -)}(\textbf{k},0)$ on the r.h.s of eq.s (\ref{Gl85}), (\ref{Gl86}) represents the lowest order term in the $\eta_\delta$-expansion of the exact solution $\phi_2^{(1, -)}(\textbf{k},t)$ at $t = 0$. In the same way one finds
\begin{equation}
\label{Gl87}
\begin{aligned}
\phi_{{\mathrm{TEO}}_1}^{(2, -)}(\textbf{k},t) \, \sim \, &[  1  - O(\eta_\delta \, \tau^{2 \delta - 1}) ] \, [ 1 + O(\eta_\delta)]  \, \phi_{{\mathrm{TEO}}_1}^{(2, -)}(\textbf{k},0) 
\\
\phi_{{\mathrm{TEO}}_2}^{(2, -)}(\textbf{k},t)   \, \sim \,
&A^{(2,-)}_2(\textbf{k}) \, e^{ i \tau} |\tau|^{ \delta - 1} \,  [1 \, + \, O(\eta_\delta \, \tau^{2 \delta - 1}) ] \, [ 1 + O(\eta_\delta)] \, \phi_{{\mathrm{TEO}}_1}^{(2, -)}(\textbf{k},0)   .
\end{aligned}
\end{equation}

The asymptotic TEO results (\ref{Gl85}) - (\ref{Gl87}) agree with the outcome of the asymptotic treatment of the approximate PT solutions, eq.s (\ref{Gl78}).
\\
\section*{VI. CONCLUSIONS}
It has been shown that for the Weyl-Dirac equation in planar Bianchi-type-I background spacetimes with power law scale factors (pBI), a simple parameter transformation (PT) exists relating all exact Weyl-spinor solutions with backgrounds in the same equivalence class $\boldsymbol{ [\,\delta \,]}$ of pBI line elements. This PT is exact, which can be proven by either employing the general exact ansatz for the time evolution operator (TEO), or a diffeomorphism restricted by an additional constraint, in combination with an appropriate Weyl-scaling. As a consequence, the knowledge of a single arbitrary member of the pertaining equivalence class of exact Weyl-spinor solutions $\boldsymbol{ [\,\bar{\delta} \,]}$ suffices to calculate all elements of that class. This has explicitly been demonstrated for the two classes $\boldsymbol{[\, \bar{1} \,]}$ and $\boldsymbol{[\, \overline{1/2} \,]}$. The exact results of $\boldsymbol{[\, \bar{0} \,]}$, $\boldsymbol{[\, \overline{\infty} \,]}$ were also listed.

The PT method is still very useful if no exact solutions are available, since it also admits the calculation of all approximate solutions of a class $\boldsymbol{ [\,\bar{\delta} \,]}$ in much the same way as in the exact case. As an illustrative example, the class $\boldsymbol{[\, \overline{1/4} \,]}$ has been treated. In particular, the approximate solutions for a background described by the anisotropic planar Kasner line element have been determined. These have been recently computed utilizing the TEO method. The comparison with the outcome of the present work shows agreement in the limiting cases $t \rightarrow 0$ and $t \rightarrow \infty$. On the other hand, there are also some differences between the two methods: The TEO approach renders approximate solutions which are only reliable to lowest order in the expansion parameter $\eta_\delta$. However, merely within the framework of the TEO formalism can the early- and late-time results be correctly combined into an aproximate solution. But due to the complicated form of the TEO it seems, apart from special cases such as $\delta = 1$ or $k_3 = 0$, impossible to get exact analytical results. In contrast, the PT formalism immediately provides, once a single (exact or approximate) solution is known, the entire corresponding equivalence class of solutions. Hence, both methods complement each other as far as the computation of approximate solutions is concerned.
\\
\section*{ACKNOWLEDGMENTS}
The author is grateful to K. H. Lotze and A. Wipf for helpful comments and support of this work.
\\
\begin{appendix}
\renewcommand{\theequation}{A\arabic{equation}}
\setcounter{equation}{0}
\section*{APPENDIX: THE LIMITING CASE $a \rightarrow 0$ } 
We first consider the general negative chirality bispinor solutions of the Minkowski class $\boldsymbol{[\, \bar{1} \,]}$, which has been given in eq. (\ref{Gl42}). For $a \rightarrow 0$ one obtains $\psi_\textbf{k} ^{' (l, -)} \rightarrow  \psi_\textbf{k} ^{' (l, -)}|_{\mathrm{FLRW}}$, with

\begin{equation}
\label{Gl43}
\psi_\textbf{k} ^{' (l, -)}(\textbf{x},t)|_{\mathrm{FLRW}}  =  d^{' (l, -)}_\textbf{k} \, e^{i\textbf{k} \textbf{x}} \,  e^{ i \, (-1)^l k \, \mathrm{sign} k_3
\, \ln \left( \frac{t}{t_A} \right) } \, t^{- \frac{3}{2}}
\left( \begin{array}{c} 
1
\\
i \,\frac{(- 1)^l \, k\, \mathrm{sign} k_3 \, - \, k_3 }{k_2 + i k_1}
\\
- \,1
\\
- i \, \frac{(- 1)^l \, k\, \mathrm{sign} k_3 \, - \, k_3 }{k_2 + i k_1}
\end{array} \right)
\end{equation}

($k \equiv \sqrt{\kappa^2 + k^2_3} \equiv \sqrt{k_1^2 + k_2^2 + k_3^2}$), which is consistent with an earlier outcome\footnote{See ref. \cite{Barut}, p. 3709; the factors $1 \pm 3 i a_0/4k$ must be replaced by unity.}.

Next, we investigate the class $\boldsymbol{[\, \overline{1/2} \,]}$. The treatment of this case gets much more involved than the previous one owing to the significantly more complicated solutions. It is convenient to take as representative the stiff-fluid element (\ref{Gl33}) with solutions $\phi^{(l, -)}(2 \textbf{k}, \sqrt{t})$, where $\phi^{(l, -)}(\textbf{k}, t)$ are given by (\ref{Gl31}), (\ref{Gl32}). The general Weyl-spinor solutions $\in \boldsymbol{[\, \overline{1/2} \,]}$ are then:

\begin{equation}
\label{Gl44}
\begin{aligned}
\phi_1^{' (1, -)}(\textbf{k}, t) \, = \,& z_a^{- 1/4} \, e^{z_a/2} \, W_{- b_a; \frac{1}{4}}(- z_a)
\\
\phi_2^{' (1, -)}(\textbf{k}, t) \, = \,&  \frac{\sqrt{- 2  i a k_3}}{i k_1 + k_2} \, z_a^{- 3/4} \, e^{ - z_a/2}  \left[ i \, \frac{\eta}{a} \, W_{-b_a; \frac{1}{4}}(- z_a) \, - \,  W_{ 1 - b_a; \frac{1}{4}}(- z_a) \right]
\\
\phi_1^{' (2, -)}(\textbf{k}, t) \, = \,& z_a^{- 1/4} \, e^{z_a/2} \, W_{b_a; \frac{1}{4}}(z_a)
\\
\phi_2^{' (2, -)}(\textbf{k}, t) \, = \,& i \, \frac{(k_2 - i k_1) \, \mathrm{sign}k_3}{\sqrt{2 i a k_3}} \, z_a^{- \frac{3}{4}} \, e^{- \frac{z_a}{2} }  \left[ W_{b_a; \frac{1}{4}}(z_a) \, + \, \frac{2 i \frac{\eta}{a} - 1}{2} \, W_{b_a - 1; \frac{1}{4}}(z_a) \right]
\end{aligned}
\end{equation}

where $z_a := - 2 i k_3 t^a/a, \,  \eta := \kappa^2/2 k_3$, and $b_a := 1/4 + i \eta/a$. The associated line element is given by (\ref{Gl10}) with $\mu = a, \, \nu = 1 - a/2$. We study first $\phi_1^{' (2, -)}$ and write it as the confluent hypergeometric function \cite{Erdelyi}:

\begin{equation}
\label{Gl45}
\phi_1^{' (2, -)}(\textbf{k}, t)  =  \Psi\left(- i \, \frac{\eta}{a}; \, \frac{1}{2}; \, z_a\right)   .
\end{equation}

Furthermore, on defining the auxiliary variable (see ref. \cite{Erdelyi}, p.281 (22))

\begin{equation}
\label{Gl46}
- i \, \zeta_a = \frac{1}{2} \, \sqrt{z_a} \, \sqrt{z_a - b_a} \, - \, b_a \, \log \frac{(\sqrt{z_a} + \sqrt{z_a - 4 b_a})^2}{4 b_a}
\end{equation}

(log denotes the principal branch of log z, and for simplicity we put from now on: $a > 0$), one gets after some algebra for $a \ll 1$, i.e. $|\zeta_a|  \gg 1$:

\begin{equation}
\label{Gl47}
\begin{aligned}
\phi_1^{' (2, -)}(\textbf{k}, t) \, = &\, \left\lbrace -  \left[1 + \frac{2 i}{a k_3} (k^2 + a k_3^2 \ln t + O(a^2))  
\right]  \right\rbrace^{- 1/4}
\\
& \times \, b_a^{b_a} \, e^{- b_a} \, e^{b_a \left( \ln \frac{k + |k_3|}{k - |k_3|}  + i \pi (2 j + 1) \right)} \, e^{z_a/2} \, e^{- \frac{k}{2 |k_3|} z_a [1 + O(a^2)]}   .
\end{aligned}
\end{equation}

Turning to $\phi_2^{' (2, -)}$ and using a recurrence relation for $W_{\alpha; \beta}$ \cite{Abramowitz} one has:

\begin{equation}
\label{Gl48}
\begin{aligned}
\phi_2^{' (2, -)}(\textbf{k}, t) = \frac{-\sqrt{- 2 i a k_3}}{ik_1 + k_2} \, \frac{e^{- z_a + (z_a)_{\widetilde{A}}}}{\sqrt{z_a}}  \Bigg[ & \frac{1 + 4 b_a - 4 z_a}{4} \, \Psi\left(\frac{1}{4} - b_a; \frac{1}{2}; z_a\right) 
\\
&+ \, \Psi \left(-\frac{3}{4}-b_a; \frac{1}{2}; z_a \right)\Bigg]
\end{aligned}
\end{equation}

($(z_a)_{\widetilde{A}} \equiv - 2 i k_3 t_{\widetilde{A}}^a/a$). With the above introduced variable $\zeta_a$ one finds eventually:

\begin{equation}
\label{Gl49}
\begin{aligned}
\phi_2^{' (2, -)}(\textbf{k}, t) \, = \, & \phi_1^{' (2, -)}(\textbf{k}, t) \, \frac{-\sqrt{- 2 i a k_3}}{ik_1 + k_2} \, \frac{e^{- z_a + (z_a)_{\widetilde{A}}}}{\sqrt{z_a}}   \, \Bigg[ \frac{1}{2} \, + \, i \, \frac{\eta}{a} \, - \, z_a \, +
\\
&+  \, \frac{(1 + b_a)^{1 + b_a} \, e^{- (1 + b_a)} \, e^{(1 + b_a) \left( \ln \frac{k + |k_3|}{k - |k_3|}  + i \pi (2 j + 1) \right)} \,}{b_a^{b_a} \, e^{- b_a} \, e^{b_a \left( \ln \frac{k + |k_3|}{k - |k_3|}  + i \pi (2 j + 1) \right)} \,} \, [1 + O(a)]  \Bigg]   .
\end{aligned}
\end{equation}

By use of eq.s (\ref{Gl47}), (\ref{Gl49}) follows then for $a \ll 1$:

\begin{equation}
\label{Gl50}
\left( \begin{array}{c} 
e^{ i P^{'}_3  } \ \phi_1^{' (2, -)}
\\
e^{- i P^{'}_3 } \ \phi_2^{' (2, -)}
\end{array}   \right)
\, = \, \phi_1^{' (2, -)} \, e^{- \frac{z_a - (z_a)_{\widetilde{A}}}{2} } \,
\left( \begin{array}{c} 
1 
\\
i \, \frac{k\, \mathrm{sign} k_3 \, - \, k_3 }{k_2 + i k_1}
\end{array}   \right) \,  [1 + O(a)]   ,
\end{equation}

and the very same procedure as above leads to the corresponding result:

\begin{equation}
\label{Gl51}
\left( \begin{array}{c} 
e^{ i P^{'}_3  } \ \phi_1^{' (1, -)}
\\
e^{- i P^{'}_3 } \ \phi_2^{' (1, -)}
\end{array}   \right)
\, = \, \phi_1^{' (1, -)} \, e^{- \frac{z_a - (z_a)_{\widetilde{A}}}{2} } \,
\left( \begin{array}{c} 
1 
\\
i \, \frac{- k\, \mathrm{sign} k_3 \, - \, k_3 }{k_2 + i k_1}
\end{array}   \right) \,  [1 + O(a)]   ,
\end{equation}

with

\begin{equation}
\label{Gl52}
\begin{aligned}
\phi_1^{' (1, -)}(\textbf{k}, t) \, =&  \ e^{i \frac{\pi}{4} \mathrm{sign} k_3} \, e^{z_a} \, \Psi\left(\frac{1}{2} +  i \, \frac{\eta}{a}; \, \frac{1}{2}; \, - z_a\right)
\\
=& \, \left\lbrace -  \left[1 + \frac{2 i}{a k_3} (k^2 + a k_3^2 \ln t + O(a^2))  
\right]  \right\rbrace^{- 1/4}
\\
& \times \, b_a^{b_a} \, e^{- b_a} \, e^{b_a \left( \ln \frac{k + |k_3|}{k - |k_3|}  + i \pi (2 j + 1) \right)} \, e^{z_a/2} \, e^{ \frac{k}{2 |k_3|} z_a [1 + O(a^2)]}   .
\end{aligned}
\end{equation}

Upon insertion of (\ref{Gl50}), (\ref{Gl51}) into (\ref{Gl25}) one recovers for $a \rightarrow 0$ again $\psi^{'}_{\mathrm{FLRW}}$ as found in (\ref{Gl43}).
\end{appendix}
\\

\end{document}